\documentclass[iop,onecolumn]{emulateapj}

\newcommand\cf{cf.}
\newcommand\eg{e.g.}
\newcommand\etc{etc.}
\newcommand\chandra{\textit{Chandra}}
\newcommand\xspec{\textsc{xspec}}
\newcommand\ciao{\textsc{ciao}}
\newcommand\sherpa{\textsc{sherpa}}
\newcommand\projct{\textsc{projct}}
\newcommand\clmass{\textsc{clmass}}
\newcommand\wabs{\textsc{wabs}}
\newcommand\mekal{\textsc{mekal}}
\newcommand\rhog{\rho_{\rm g}}
\newcommand\ith{$i^{\rm th}$}
\newcommand\jth{$j^{\rm th}$}
\newcommand\nth{$n^{\rm th}$}
\newcommand\mh{m_{\rm H}}
\newcommand\ie{i.e.}
\newcommand\emiss[2]{{\rm EM}_{{#1},{#2}}}
\newcommand\nelec[1]{n_{{\rm e}#1}}
\newcommand\nh{n_{\rm H}}
\newcommand\rhobar{\overline{\rho}}
\newcommand\msun{M_\odot}
\newcommand\rover{r_{500}}
\newcommand\rhover{r_{2500}}

\shorttitle{Model-independent X-ray mass determinations}
\shortauthors{Nulsen, Powell \& Vikhlinin}

\begin{document}


\title{Model-independent X-ray mass determinations}


\author{P. E. J. Nulsen\altaffilmark{1,4}, S. L. Powell\altaffilmark{1,2,3}
  and A. Vikhlinin\altaffilmark{1}} 


\altaffiltext{1}{Harvard-Smithsonian Center for Astrophysics, 60
  Garden St, Cambridge, MA 02138}
\altaffiltext{2}{University of Southampton, Highfield, Southampton,
  SO17 1BJ, UK}
\altaffiltext{3}{Institute of Astronomy, Madingley Road, Cambridge,
  CB3 0HA, UK} 
\altaffiltext{4}{email: pnulsen@cfa.harvard.edu}


\begin{abstract}
A new method is introduced for making X-ray mass determinations of
spherical clusters of galaxies.  Treating the distribution of
gravitating matter as piecewise constant and the cluster atmosphere
as piecewise isothermal, X-ray spectra of a hydrostatic atmosphere are
determined up to a single overall normalizing factor.  In contrast to
more conventional approaches, this method relies on the minimum of
assumptions, apart from the conditions of hydrostatic equilibrium and
spherical symmetry.  The method has been implemented as an \xspec{}
mixing model called \clmass{}, which was used to determine masses for
a sample of nine relaxed X-ray clusters.  Compared to conventional
mass determinations, \clmass{} provides weak constraints on values of
$M_{500}$, reflecting the quality of current X-ray data for cluster
regions beyond $\rover$.  At smaller radii, where there are high
quality X-ray spectra inside and outside the radius of interest to
constrain the mass, \clmass{} gives confidence ranges for $M_{2500}$
that are only moderately less restrictive than those from more
familiar mass determination methods.  The \clmass{} model provides
some advantages over other methods and should prove useful for mass
determinations in regions where there are high quality X-ray data.
\end{abstract}





\section{Introduction}

The need for accurate masses for galaxy clusters is motivated by more
than their significance as fundamental parameters of the most massive,
dynamically stable structures in the universe.  Because gravity is
dominant on the largest scales in the universe, masses of the largest
virialized structures are important cosmological probes
\citep[\eg][]{v05,ars08,sr09}.  For example, the growth of structure
at high mass scales provides a sensitive means to measure the power
spectrum of density fluctuations that emerged from the early universe
and to determine the time dependence of the rate of cosmic expansion,
hence placing useful constraints on cosmological models
\citep{vkb09}.

The dominance of gravity in clusters also means that there are several
avenues for clusters mass determinations, which can be classified
broadly as dynamical \citep[\eg][]{ggm98,rgk03}, gas dynamical
\citep[\eg][]{pap05,vkf06} and lensing \citep[\eg][]{sks05,lrj07}.  No
approach is free of weaknesses.  Dynamical mass determinations are
limited by our inability to measure velocities transverse to the
line-of-sight \citep[\eg][]{mb09}.  Gas dynamical methods rely on the
intracluster medium (ICM) being hydrostatic \citep[\cf][]{mv07} and
they require accurate determinations of the total pressure of the ICM
\citep[\eg][]{cfv08,mhb08}.  Strong lensing is confined to the highest
density regions of rich clusters \citep[\eg][]{nte09}, while weak
lensing is affected by substantial statistical uncertainties for
individual clusters, line-of-sight projections and the mass-sheet
degeneracy \citep[\eg][]{mwl01,h03,bls04,zfb08}.  Many of these issues
can be addressed by judicious choice of targets, but the most reliable
results are likely to come from the application of multiple mass
determination methods to cross check and augment one another
\citep[\eg][]{mhb07,cfv08,nte09}.

This paper introduces a new approach to X-ray mass determinations.
The method relies on the same basic assumptions as other X-ray
mass determinations, \ie, that the intracluster gas is hydrostatic and
spherically symmetric.  However, it largely avoids the additional,
model-dependent assumptions that are required by other approaches.
The method is outlined in section \ref{sec:method} and applied to a
sample of clusters with high quality \chandra\ data in section
\ref{sec:results}.  Strengths and weaknesses of the method are related
to the results in section \ref{sec:discussion}.

Following \citet{vkf06}, all distances are computed assuming a flat
$\Lambda$CDM cosmology, with $\Omega_{\rm M} = 0.3$ and a Hubble
constant of $72\rm\ km\ s^{-1}\ Mpc^{-1}$.  For each cluster, the
radii $\rover$ and $\rhover$ enclose regions with densities of 500 and
2500, respectively, times the critical density at the cluster
redshift, while $M_{500}$ and $M_{2500}$ are the corresponding
enclosed masses.  Confidence ranges are determined at the 90\% level
except where stated otherwise.

\section{Method} \label{sec:method}

X-ray mass determinations rely on the ability of X-ray observations to
determine the temperature and density of the hot ICM.  In an ICM
dominated by hot gas, the pressure is determined by its temperature
and density.  Undetected components of the ICM, \ie, nonthermal
particles, magnetic fields and turbulence, can add to the effective
pressure, although simulations \citep[\eg][]{nvk07} and observations
\citep[\eg][]{cfv08} suggest that these contributions are modest, of
the order of $10\%$, in relaxed appearing clusters.  Since the purpose
of this paper is to introduce a new approach to X-ray mass
determinations, such effects are not considered further here, but they
need to be understood for precise mass determinations.

If the gravitational potential is spherically symmetric and the ICM is
in hydrostatic equilibrium, the distribution of the ICM is governed by
the equation of hydrostatic equilibrium,
\begin{equation} \label{eq:static}
{dp \over dr} = - \rhog(r) {G M(r) \over r^2}, \label{eqn:hs}
\end{equation}
where $p(r)$ is the gas pressure, $\rhog(r)$ is its density and $M(r)$
is the total gravitating mass within the radius, $r$.  If the pressure
and density distributions of the ICM are known, $M(r)$ can be
determined from this equation.  In practice, properties of the ICM
(particularly its temperature) are only determined at discrete radii,
so that the pressure derivative is constrained poorly, if at all, by
the data.  With few exceptions \citep{nb95,aba04}, mass determinations
circumvent this difficulty by employing model-dependent assumptions to
supplement equation (\ref{eqn:hs}).  Details vary widely, but typical
approaches assume analytic forms, with a modest number of parameters,
either for the mass profile \citep[\eg][]{mhb07}, or for the
temperature and density profiles \citep[\eg][]{vkf06}.  For example,
using the latter approach, substituting the analytic form for the
pressure and density profiles in equation (\ref{eqn:hs}) gives an
analytic expression for $(1/\rhog) dp/dr$, with parameters that can be
determined by fitting the analytic temperature and density profiles to
the discrete temperatures and densities obtained from X-ray spectra.

Also with few exceptions \citep[\eg][]{mhb07}, several steps are
required to obtain a cluster mass, so that a confidence range for the
mass is determined indirectly from fitting the X-ray spectra.  For
example, data may be ``deprojected'' to obtain gas temperatures and
densities (with strongly correlated errors).  Deprojected temperatures
and densities are then fitted with analytic models to determine their
parameters, which are used to obtain the mass, as described above.
Doing mass determinations in several steps simplifies the overall
process, but it complicates error determinations, since errors must be
propagated through all steps.  While there are well-known statistical
methods for computing confidence ranges in such cases, the many
parameters involved (at the very least, one temperature and one
density for each X-ray spectrum), make the process cumbersome and
expensive.  Error propagation is more flexible when it is possible to
marginalize over fitting parameters directly to compute a confidence
range for the mass.  This is considerably easier when the mass can be
determined directly in terms of parameters used to fit to the X-ray
data, as in the approach used here.

\subsection{A model-independent approach} \label{sec:outline}

A simple model can be used to approximate any spherical cluster and
its atmosphere well.  The cluster volume is divided into a series of
spherical shells centered on the cluster and, in addition to the
assumptions of spherical symmetry and hydrostatic equilibrium, it is
assumed that
\begin{list}{}{\setlength{\topsep}{3pt}\setlength{\parskip}{0pt}
    \setlength{\parsep}{0pt}\setlength{\itemsep}{3pt}}
\item[1)] the gravitating matter density in each spherical shell is
  constant and
\item[2)] the gas in each spherical shell is isothermal.
\end{list}
Assuming that the gravitating matter density is constant in shells
provides, essentially, the simplest continuous mass distribution that
is nonsingular at $r = 0$, which can approximate any real mass
distribution well with a large number of shells.  Similarly, assuming
that the shells are isothermal results in, perhaps, the simplest gas
model that can approximate any real, spherically symmetric, gas
temperature distribution well with a large enough number of shells.
Projected onto the sky, the spectrum for any region of the model
cluster is a sum of optically thin, thermal models, a form that can be
expressed conveniently as an \xspec{} mixing model \citep{a96}.

Although the term is inappropriate in the strict mathematical sense,
since this model can approximate arbitrary mass and temperature
distributions, we call the method ``model-independent.''  The model
used here is one instance of a broad class of (model-independent and
model-dependent) approximations that could be used for the gravitating
matter distribution (alternative models would differ in equations
\ref{eqn:mass} and \ref{eqn:dphi} below).

Denoting radii measured from the center of the cluster by $r$ and the
shell boundaries by $r_1, r_2, \ldots, r_{n+1}$ in increasing order, the
\ith{} shell is the region with $r_i < r < r_{i+1}$.  If the
gravitating matter density within the \ith{} shell is $\rho_i$, then
the gravitating mass distribution within the shell is given by
\begin{equation}
M(r) = M_i + {4\pi\over 3} \rho_i (r^3 - r_i^3), \label{eqn:mass}
\end{equation}
where $M_i = M(r_i)$ is the total gravitating mass within $r_i$.  The
gravitational potential within the \ith{} shell is then given by 
\begin{equation}
\Phi(r) - \Phi(r_i) = \int_{r_i}^r {G M(r) \over r^2} \, dr
= \left[ {GM_i \over r_i} + {2\pi\over3} G \rho_i (r - r_i) (r + 2
  r_i) \right] {r - r_i \over r}.   \label{eqn:dphi}
\end{equation}

Since the gas in each shell is assumed to be isothermal and
hydrostatic, within the \ith{} shell the electron density can be
expressed as
\begin{equation}
\nelec{} (r) = \nelec{,i} \exp \left[ - {\mu \mh \{ \Phi(r) -
    \Phi(r_i) \} \over kT_i} \right], 
\qquad {\rm for}\ r_i < r < r_{i+1}, \label{eqn:ne} 
\end{equation} 
where $\nelec{,i}$ is the electron density at the inner edge of
the \ith{} shell, $T_i$ is the gas temperature in that shell, $\Phi(r)$
is the gravitational potential, $k$ is Boltzmann's constant and
$\mu\mh$ is the mean mass per particle in the gas.  For the gas to be
hydrostatic, the pressure must be continuous from shell to shell, but,
as the temperature is generally discontinuous at shell boundaries, so
is the gas density.  Pressure continuity between shells $i$ and $i+1$
requires 
\begin{equation}
\nelec{,i + 1} T_{i+1} = \nelec{,i} \exp \left[ - {\mu \mh \{
    \Phi(r_{i + 1}) - \Phi(r_i) \} \over kT_i} \right] T_i, \label{eqn:cont} 
\end{equation}
which determines $\nelec{,i + 1}$ in terms of $\nelec{,i}$, the
temperatures and the potential.

Together, these results show that when the gravitating matter
density and gas temperature are specified for each shell, under the
model assumptions, the gas density is determined throughout the model
atmosphere, up to a single scale factor that may be regarded as the
electron density, $\nelec{,1}$, at the inner edge of the innermost
shell (usually $r=0$).  

\begin{figure}[t]
\centerline{\includegraphics[width=0.6\textwidth]{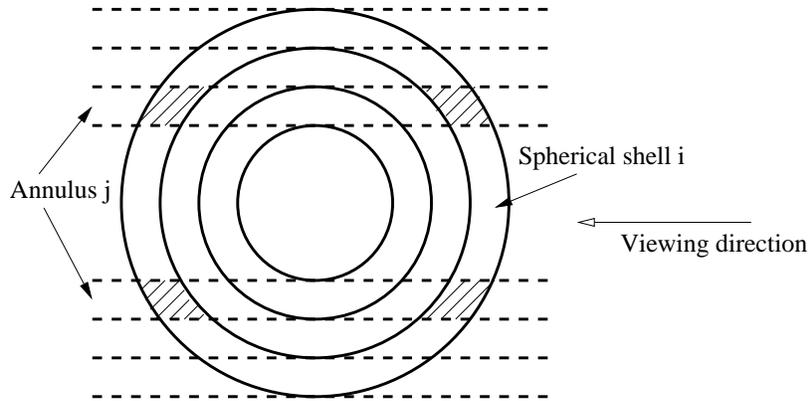}}
\caption{Shells and annuli used for the mass model.  This schematic
  shows a cross-section through a cluster center that contains our
  line-of-sight.  Dashed lines show boundaries of cylindrical regions
  that project onto the annuli on the sky from which X-ray spectra are
  extracted.  Full lines show boundaries of the corresponding
  spherical shells.} \label{fig:annuli}
\end{figure}

When observing a spherical cluster, the normal practice is to divide
the sky into concentric annuli centered on the cluster and extract a
spectrum from each annulus.  An annulus is the projection onto the sky
of a cylindrical region in the cluster.  The schematic in
Fig.~\ref{fig:annuli} shows sections through some cylinders (bounded
by dashed lines) in a plane that contains our line-of-sight through
the cluster center, \ie, the common axis of the cylinders.  We now
define the spherical shells discussed above to have radial bounds
matching those of the cylinders (full lines in Fig.~\ref{fig:annuli}).
Denoting cylindrical radii, measured from the line-of-sight through
the center of the cluster, by $\varpi$, the \jth{} cylindrical region
is then defined by $\varpi_{j} \le \varpi \le \varpi_{j + 1}$ and, by
construction, $r_i = \varpi_i$, which simplifies some of the following
results.  The same geometric arrangement is used for deprojections.

To complete the spectral model, we need to calculate the emission
measure of the gas from the \ith{} shell that appears in the \jth{}
annulus, \ie, the emission measure of the gas in the intersection
between the \ith{} shell and the \jth{} cylinder (hatched region in
Fig.~\ref{fig:annuli}),
\begin{equation}
\emiss{i}{j} = \int_{\rm Intersection} \nelec{} \nh \, dV 
= 4\pi \int_{r_i}^{r_{i+1}} \nelec{}(r) \nh(r) \left( \sqrt{r^2 -
  r_j^2} - \sqrt{r^2 - r_{j + 1}^2} \right) r\, dr, \label{eqn:em}
\end{equation}
where each square root should be taken as zero if its argument is
negative.  These integrations are carried out numerically.  Note that
for $i < j$, since the \ith{} spherical shell does not intersect the
\jth{} cylinder, $\emiss{i}{j} = 0$.  The model spectrum for the
\jth{} annulus is given as a sum over the shells of the thermal model
for each spherical shell, weighted by its emission measure in the
\jth{} cylinder, \ie, the spectrum is
\begin{equation}
F_j (E) = \sum_{i = j}^n \emiss{i}{j} f(E, T_i, A_i),  \label{eqn:spec}
\end{equation}
where $f(E, T_i, A_i)$ is the thermal spectrum for the \ith{}
spherical shell (for unit emission measure).  The parameter, $A_i$, is
a placeholder for all thermal model parameters, including abundances,
apart from the temperature.

High quality spectra are required to obtain good results when fitting
this model.  Since the demands of the model are similar to those of
deprojections, it is expected that the minimum required source count
per spectrum is several thousand.  There is a clear trade off between
reducing shell width to gain spatial resolution and increasing it to
improve spectral signal-to-noise (see section \ref{sec:strengths}).

\subsection{Implementation} \label{sec:implement}

The model outlined above has been implemented as an \xspec{} mixing
model called \clmass.  This mixing model has two parameters per
spherical shell, the gas temperature and the gravitating matter density.
The temperature for each shell must be linked (equated) to the
temperature parameter of a thermal model for that shell.  As noted,
there is a single overall gas density scale that is free in the model.
For the mixing model, it is convenient to make the \xspec{} norm for
the innermost shell the free parameter instead of the central gas
density, $\nelec{,1}$.  The \xspec{} norm is the emission measure
multiplied by a scale factor that is the same for each shell and
depends on the distance to a cluster.  In using the \clmass{} model, the
\xspec{} norm for every outer shell is linked (equated) to that of the
central shell and the X-ray flux for the \ith{} annulus (equation
\ref{eqn:spec}) is given by
\begin{equation}
S_j (E) = K \sum_{i = j}^n W_{i,j} f(E, T_i, A_i),  \label{eqn:flux}
\end{equation}
where $K$ is the single free norm (for the innermost shell) and there
are $n$ spherical shells.  The weights, $W_{i,j} = \emiss{i}{j} /
\emiss11$, are completely determined by the parameters of the
mixing model.


Frequently, a cluster extends beyond the field of view of an X-ray
detector, so that there is X-ray emission in the field of view
originating from gas lying beyond the outer edge of the outermost
spherical shell (the \nth{} shell) of the \clmass{} model.  There is no
entirely satisfactory way to deal with such emission, but we often
want to use such nonideal data for mass determinations.  Using a local
X-ray background measured immediately outside the \nth{} annulus
generally overestimates the ``background'' gas emission in all inner
shells.  Based on empirical fits to gas density profiles, the \clmass{}
model allows the alternative of assuming that the gas density profile
beyond the inner edge of the \nth{} shell follows a beta model,
\begin{equation}
\nelec{} (r) = \nelec{,0} (1 + r^2/a^2)^{-3\beta/2}, \label{eqn:beta}
\end{equation}
where $\beta$ and $a$ are constant parameters.  The gas density at the
inner edge of the \nth{} shell is still determined by pressure
continuity (equation \ref{eqn:cont}), but, instead of using the gas
density profile from equation (\ref{eqn:ne}), the emission measure for
the \nth{} shell in each annulus is determined by integrating the beta
model to infinity along our line-of-sight (giving results in terms of
incomplete beta functions).  Together with a switch that determines
whether or not this feature is used, the beta model adds three fixed
parameters to the \clmass{} model.  This feature is at odds with the
model-independent approach of \clmass{} otherwise.  Ideally, it should
be avoided by using X-ray data that cover the whole of a cluster.

For the sake of brevity, the new mass model will be referred to as
\clmass{} in the remaining discussion.  Much of that discussion would
apply to any implementation of the model, although there are some
comments on limitations that are specific to the \xspec{}
implementation, as noted.

\section{Masses for a sample of galaxy clusters} \label{sec:results}

\subsection{Sample and data preparation} \label{sec:prep}

The \clmass{} model has been applied to \chandra{} data for a
subsample of the clusters analyzed by \citet{vkf06} and \cite{vbe09}.
Target clusters were chosen to sample a broad range of cluster
temperatures, hence masses.  Clusters analyzed here are listed with
some of their basic properties in Table~\ref{tab:sample}.

\begin{table}[t]
\caption{The cluster sample} \label{tab:sample}
\begin{center}
\begin{tabular}{lccccc} 
\hline \hline
\noalign{\smallskip}
Cluster & z & $\rhover$\tablenotemark{1} & $\rover$\tablenotemark{1}
    & \chandra{} ObsID \\
& & (kpc) & (kpc) \\
\noalign{\smallskip}
\hline
\noalign{\smallskip}
A907 & 0.1603 & 501 & 1095 & 3185, 3205, 535 \\
A1413 & 0.1429 & 559 & 1300 & 1661, 5002, 5003 \\
A1991 & 0.0592 & 341 & 734 & 3193 \\
A2029 & 0.0779 & 642 & 1359 & 891, 4977, 6101 \\
A2390 & 0.2302 & 561 & 1414 & 4193 \\
A1835 & 0.2520 & 673 & 1475 & 6880, 6881, 7370 \\
A1650 & 0.0845 & 515 & 1128 & 5822, 5823, 6356, 6357,
    6358, 7242 \\
A3112 & 0.0761 & 459 & 1025 & 2216, 2516, 6972, 7323, 7324 \\
A2107 & 0.0418 & 416 & 919\tablenotemark{2} & 4960 \\
\noalign{\smallskip}
\hline
\end{tabular}
\tablenotetext{1}{Overdensity radii from \citet{vkf06}.}
\tablenotetext{2}{\chandra{} data for A2107 do not extend to
  $\rover$, so that $M_{500}$ cannot be determined for A2107.}
\end{center}
\end{table}

To simplify comparison with the results of \citeauthor{vkf06}, our
analysis parallels theirs closely, using the same X-ray spectra as far
as possible with the \clmass{} model.  Details of the spectral
reductions can be found in \citet{vmm05}.  The spectra were extracted
from annuli with radial bounds having $\varpi_{j+1} = 1.5 \varpi_j$,
for $j > 1$.  Since the gravitating matter density is constant in the
corresponding spherical shells, the matter density profile is sampled
rather coarsely in the resulting model.  To make our masses directly
comparable, we have also adopted the values of $\rhover$ and $\rover$
from \citeauthor{vkf06} (Table~\ref{tab:sample}).  Note that this
eliminates a source of uncertainty in $M_{2500}$ and $M_{500}$,
reducing confidence ranges compared to ab initio determinations of these
masses.

\citeauthor{vkf06} combined spectra from the different front
illuminated (FI) chips into single spectra.  However, because they
have significantly different responses, FI spectra were kept separate
from those extracted from the back illuminated (BI) chips.  This means
that the number of spectra per annulus for a cluster can vary in the
transition regions between FI and BI chips.  The version of \clmass{}
used here inherited from \projct{} a requirement that the same number
of spectra be used for every data group.  Particularly for
observations with ACIS-S at the aim point, this means that we were
unable to use every available spectrum for some annuli.  In cases
where a choice had to be made, we opted for the spectrum with the
greater source count.  As a result, we used slightly fewer spectra
than \citeauthor{vkf06} for some mass determinations.  This
shortcoming of \clmass{} will be remedied in a future
version.\footnote{A new version of the \xspec{} model addressing this
  and some other issues has been completed.  The new version also
  makes it easier to use alternative forms for the gravitational
  potential and a model using the Navarro, Frenk \& White potential
  has been added as a demonstration.  Verions of these models are also
  available for \ciao{} \sherpa{}.} 

A limitation of \xspec{} mixing models is that they must be applied
after all other model components.  This is inconvenient for several
reasons.  Photoelectric absorption due to foreground gas should be
applied to the spectrum emerging from each annulus, after it is
emitted by a cluster.  In practice, the foreground absorption usually
does not vary significantly across a cluster, so that it is
satisfactory to apply the same absorption model to each thermal model
before the mixing model.  However, there are clusters where the
absorbing column varies significantly, which cannot be modeled well by
the \clmass{} model in \xspec{}.

Another inconvenience arises in background corrections.  In the
standard approach, background spectra are constructed from source free
exposures, processed and scaled to match the observations.  Variations
of the particle background and the soft X-ray background with time and
position can leave residual backgrounds requiring further correction.
These are often addressed by model components added to the annular
spectra \citep{vmm05}, but this approach cannot be used in conjunction
with an \xspec{} mixing model.\footnote{In \xspec{}12, such components
  can be added by defining additional models.}  Here, the residual
backgrounds are corrected using a specially prepared \textsc{corrfile}
for each annular spectrum.  Depending on need, one or two soft thermal
components were fitted to the spectrum from the outermost annulus for
each cluster (to account for a residual soft background and any
emission from the North Polar Spur).  These components were scaled by
area and added to the existing \textsc{corrfile} for each annulus
\citep[already employed to correct for the read out
  artifact;][]{vmm05}.

Gratifyingly, inclusion of the beta model feature described in section
\ref{sec:implement} had little effect on the quality of the model fits
and on the masses.  This is to be expected, since the clusters are
barely detected in the largest annuli, so there should be little need
to account for cluster emission from regions beyond the outermost
shells.

\subsection{Cluster Masses}

\begin{figure}[t]
\centerline{\includegraphics[width=0.4\textwidth]{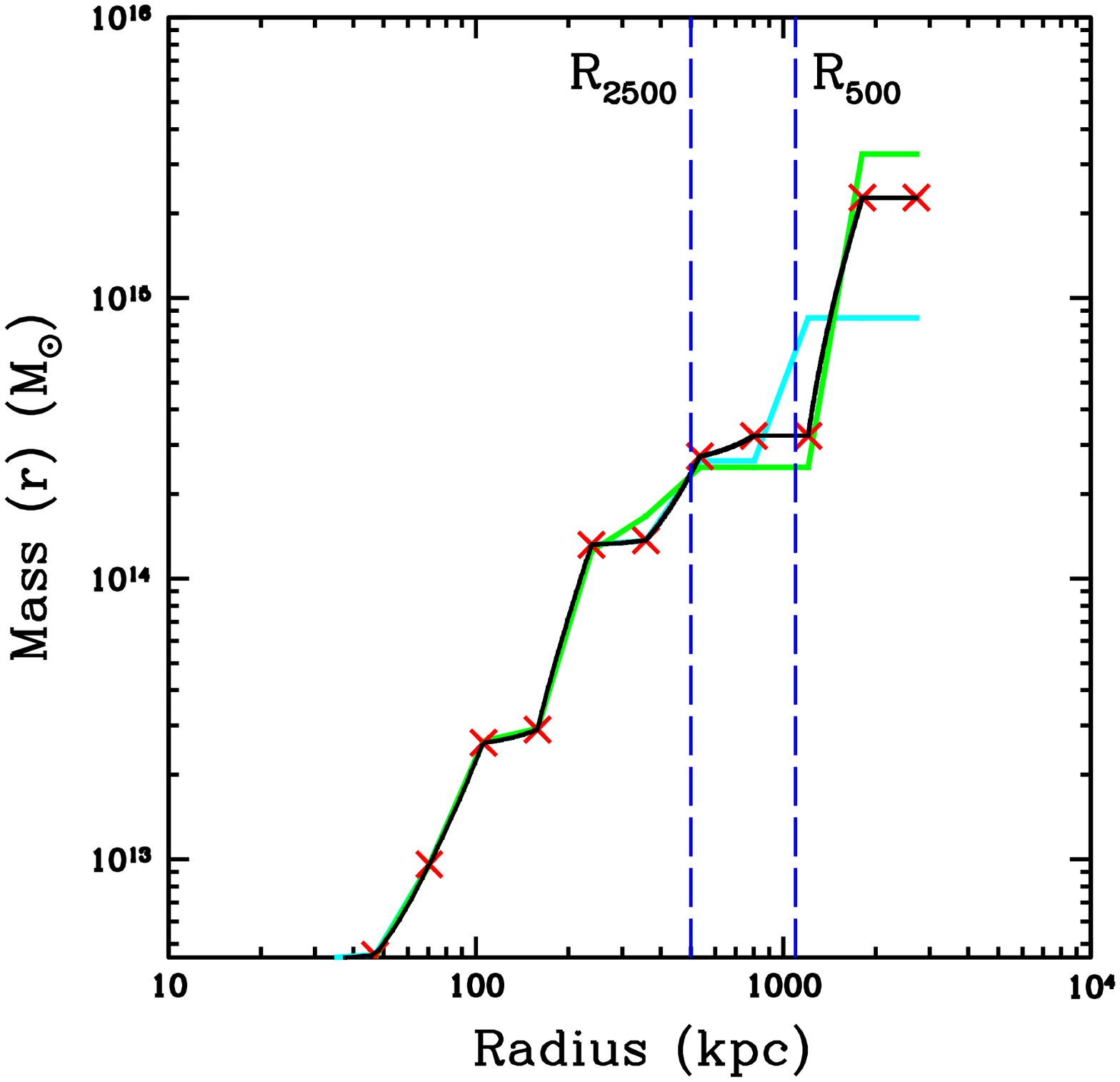}
\hbox to 0.1\textwidth{\hfil}
\includegraphics[width=0.4\textwidth]{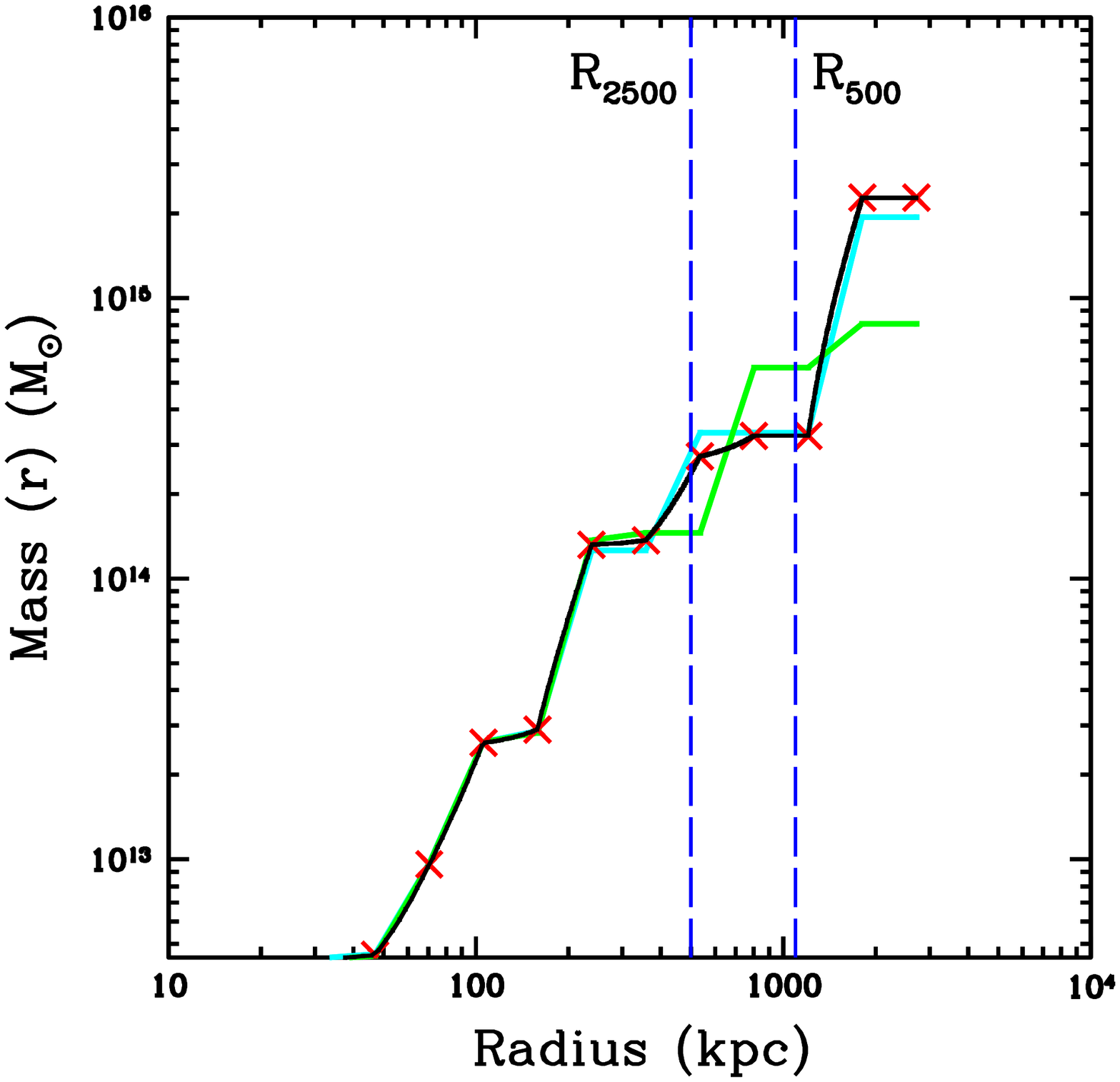}}
\caption{Mass profiles for Abell~907.  The black line shows the
  cumulative mass distribution for the best fitting \clmass{} model
  when no constraints are applied to the gravitating matter densities.
  Crosses mark the shell boundaries.  Large variations in slope from
  shell-to-shell reflect large variations in the gravitating matter
  density.  {\it Left panel:} The cyan line shows the mass profile at
  the upper 90\% confidence limit for the mass within $\rover$, \ie,
  $M_{500}$, and the green line shows the corresponding mass profile
  at the lower 90\% confidence limit.  {\it Right panel:} Same for
  $M_{2500}$.} \label{fig:a907mass}
\end{figure}

The \clmass{} mixing model was combined with an absorbed thermal
model, $\wabs{} * \mekal$, to fit the sample clusters.  Column
densities for photoelectric absorption were set to the values used by
\citet{vmm05}.  As discussed in section \ref{sec:implement}, many
parameters must be linked to use the \clmass{} model.  Since the
number of shells provided by the model is fixed, usually there are
unused parameters that must also be frozen.  Parameters were given
reasonable initial values to forestall numerical problems (section
\ref{sec:implement}).  With $\sim10$ annuli per cluster, in order to
avoid error prone manipulations of numerous parameters, these steps
were automated in \xspec{} Tcl scripts.  Abundances were generally
free in the fits, but where they were poorly constrained or took on
unreasonable values, abundances for groups of adjacent shells were
tied together to produce physically reasonable results.  Since they
have little impact on the masses, we do not consider abundances
further.  All free parameters and links used for the fits were
maintained when determining confidence ranges for the masses.

\clearpage

Initial attempts to fit the data showed that the \clmass{} model does
not generally constrain the gravitating matter density tightly for
individual shells.  This is because the gravitational potential is
governed by the accumulated mass within any radius and variations of
the gas density depend on changes in the potential.  Since there are
two integrations between the mass density and changes in the
potential, the gas density, hence the spectra, are only affected at
second order by local changes in the gravitating matter density
(equation \ref{eqn:dphi}).  Splitting the gravitating matter from one
shell into the two adjacent shells, in suitable proportions, has
little impact on the fit.  In the presence of noise, this freedom to
exchange mass from shell to shell results in large shell to shell
variations of gravitating mass density, producing noisy best fitting
mass profiles like that for Abell~907 in Fig.~\ref{fig:a907mass}.

\begin{figure}[t]
\centerline{\includegraphics[width=0.4\textwidth]{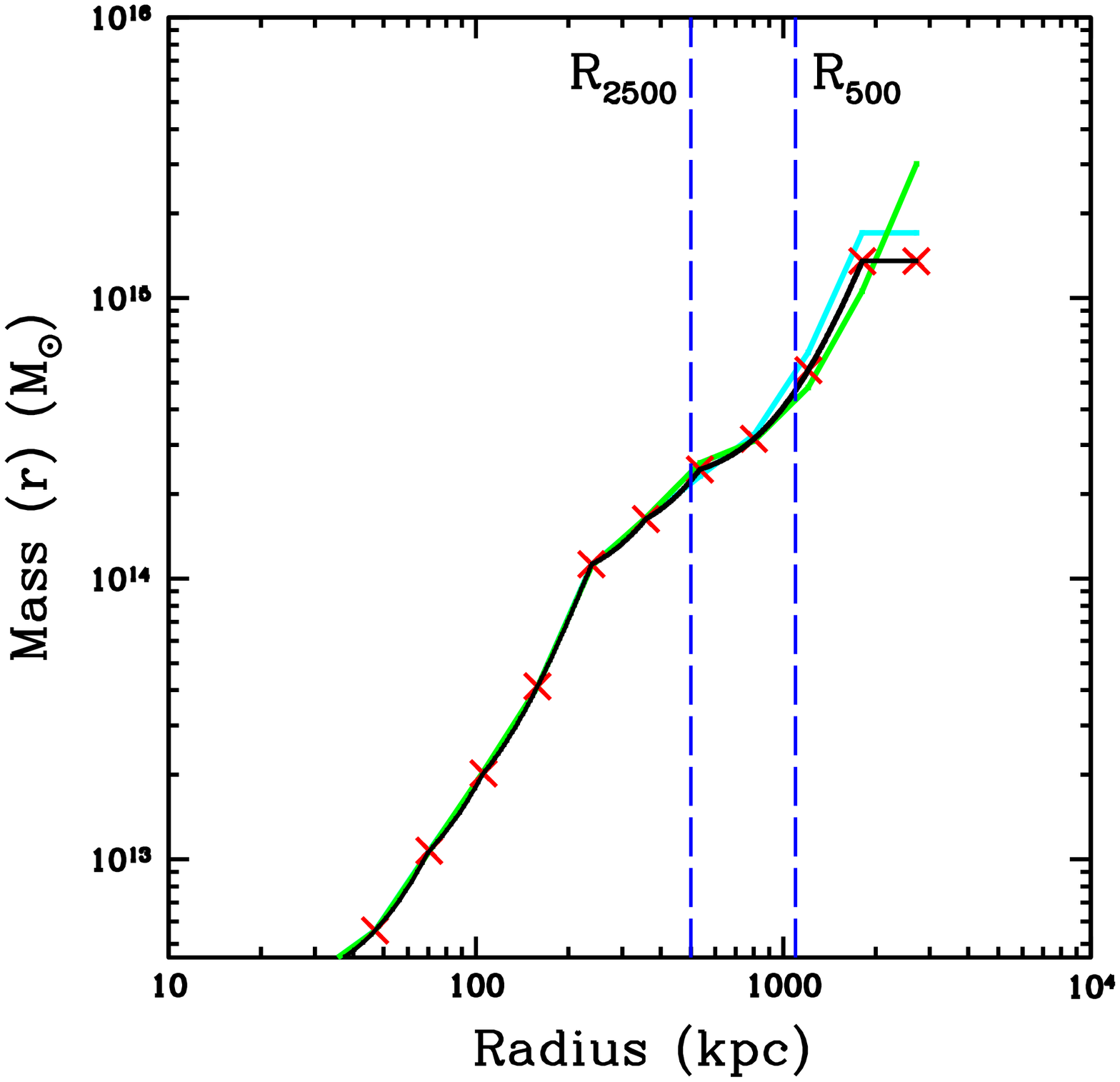}
\hbox to 0.1\textwidth{\hfil}
\includegraphics[width=0.4\textwidth]{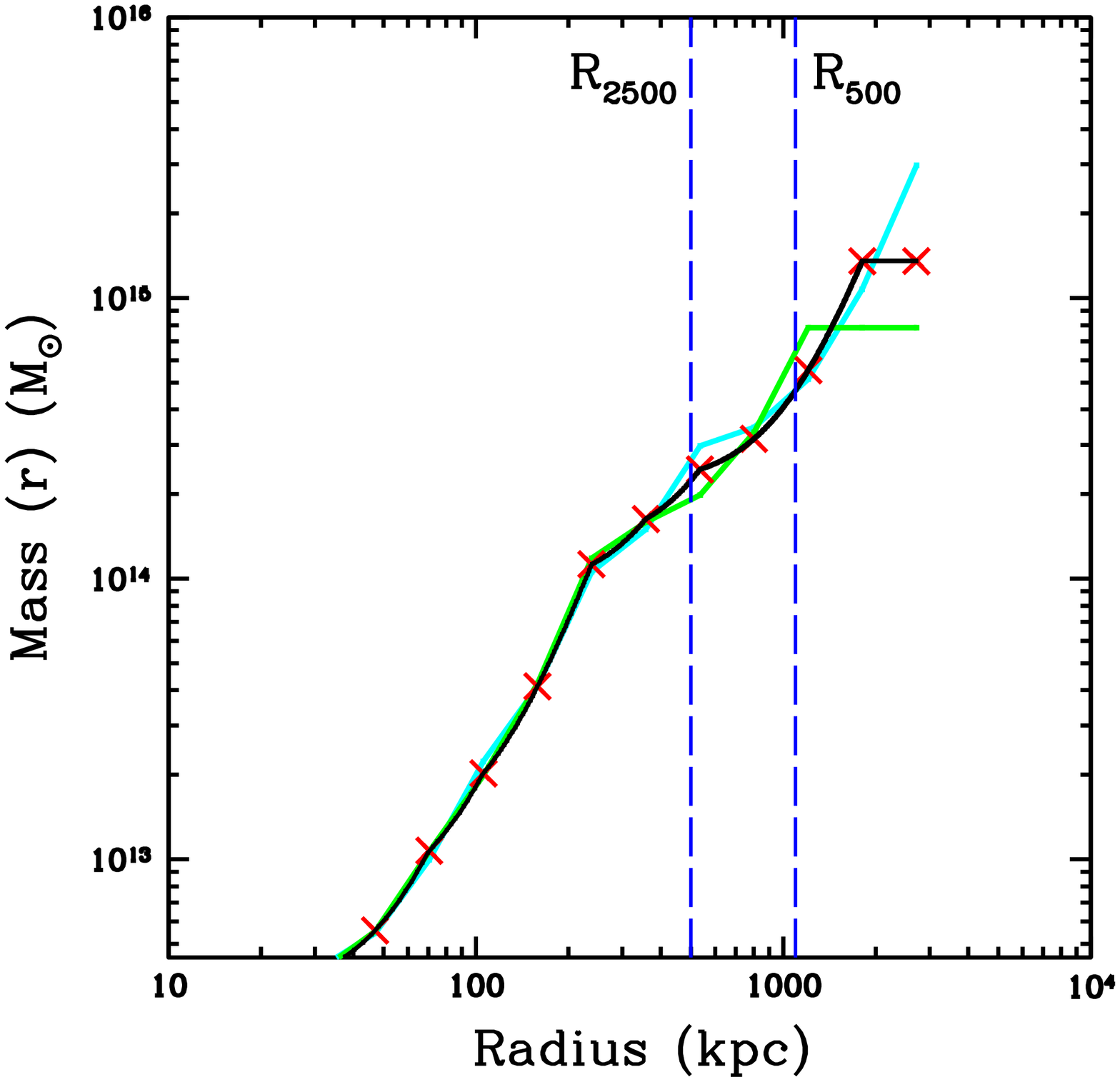}}
\caption{Mass profiles for Abell~907 with the gravitating matter
  density constrained to be monotonic.  Details as in
  Fig.~\ref{fig:a907mass}.} \label{fig:a907cons} 
\end{figure}

The freedom of gravitating mass to shift between adjacent shells can
be limited to some extent by constraining the gravitating matter
density to be a monotonically decreasing function of the radius.  This
constraint improves model behaviour with a relatively minor impact on
the goodness of fit (Table \ref{tab:mass}).  In practice, the
constraint requires gravitating mass densities to be linked for groups
of adjacent shells where the unconstrained density would otherwise
increase with radius.  Mass profiles for Abell~907, with the
gravitating matter density forced to be monotonic, are shown in
Fig.~\ref{fig:a907cons}.  Evidently, the monotonic constraint makes
the matter density profile considerably smoother.  Although this
condition is not required of the gravitating matter density
distribution, it is simple, physically reasonably and relatively weak,
while it reduces the noise in the mass profiles significantly.  As a
consequence, we prefer results for ``monotonic'' mass profiles in the
following, although some results for unconstrained mass profiles are
also given for comparison.

\begin{figure}[t]
\centerline{\includegraphics[width=0.4\textwidth]{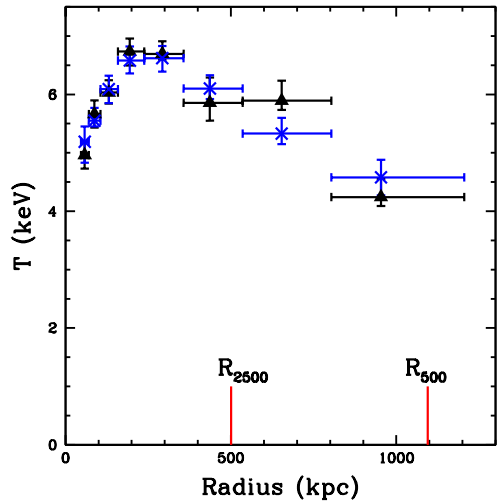}}
\caption{Temperature profiles for Abell~907.  Black triangles show
  temperatures for the best fitting ``monotonic'' \clmass{} model.
  Blue crosses show the (projected) temperature profile of
  \citet{vmm05}.  Temperature errors are 68\% confidence ranges for
  individual shells.  Radial error bars show extents of the annuli.
  \clmass{} temperatures are deprojected, which generally makes them
  noisier than projected temperatures.} \label{fig:a907temp}
\end{figure}

Once the temperatures have been determined, gas densities are governed
by the distribution of gravitating matter.  Loosely speaking, when
fitting the \clmass{} model, temperatures are determined by spectral
shapes, so that the gravitating matter densities are determined by the
radial distribution of the X-ray emission (\ie, emission measure).
However, this is only a rule of thumb.  Especially in the outskirts of
clusters where the X-ray emission is not well detected above the
background, fitted temperatures may be poorly constrained by the
spectra, so that the temperatures can also respond to variations in
X-ray brightness.  This may result in large temperature excursions in
response to poorly modeled fluctuations in X-ray brightness.  Also,
since the temperatures apply to gas in the spherical shells, they are
deprojected temperatures.  Like other deprojected temperatures, the
spectral models are coupled from shell-to-shell, making the
temperatures \etc, prone to fluctuate from shell-to-shell.  This
effect can be seen in Fig.~\ref{fig:a907temp}, which compares the best
fitting \clmass{} temperature profile for Abell~907 to the projected
temperature profile of \citet{vmm05}.

\begin{table}[t]
\caption{Cluster masses}\label{tab:mass}
\begin{center}
\begin{tabular}{lccccc}
\hline\hline
\noalign{\smallskip}
&\multicolumn{2}{c}{Unconstrained}&\multicolumn{3}{c}{Monotonic}\\
Cluster & $\chi^2/\rm dof$ & $M_{500}$ & $\chi^2/\rm dof$ & $M_{500}$
& $M_{2500}$ \\
&& ($10^{14}\ \msun$) && ($10^{14}\ \msun$) & ($10^{14}\ \msun$) \\
\noalign{\smallskip}
\hline
\noalign{\smallskip}
A907 & $692.4/626$ & $3.2_{-0.7}^{+3.2}$ & $693.7/630$ &
$4.7_{-0.5}^{+0.6}$ & $2.2_{-0.3}^{+0.4}$ \\
A2390 & $840.9/626$ & $23.0_{-14.8}^{+13.3}$ & $846.2/632$ &
$14.9_{-5.3}^{+9.1}$ & $4.0_{-1.1}^{+0.5}$ \\
A1835 & $489.4/290$ & $7.1_{-1.3}^{+17.4}$ & $491.7/293$ &
$7.1_{-1.3}^{+12.0}$ & $7.1_{-2.1}^{+1.0}$ \\
A1650 & $669.1/301$ & $5.1_{-1.5}^{+7.1}$ & $673.2/306$ &
$5.8_{-1.3}^{+3.9}$ & $2.1_{-0.2}^{+0.2}$ \\
A3112 & $545.8/314$ & $3.0_{-1.1}^{+4.4}$ & $546.7/317$ &
$2.8_{-0.9}^{+2.4}$ & $1.7_{-0.3}^{+0.3}$ \\
A2029 & $2988.0/664$ & $6.6_{-1.5}^{+4.5}$ & $2993.5/669$ &
$7.1_{-1.8}^{+4.0}$ & $5.0_{-0.6}^{+0.3}$ \\
A1991 & & & $563.7/326$ & $5.9_{-4.0}^{+4.6}$ & $0.8_{-0.2}^{+0.4}$ \\
A1413 & $449.0/313$ & $7.9_{-3.9}^{+10.5}$ & $450.3/321$ &
$9.1_{-3.2}^{+5.9}$ & $3.6_{-0.6}^{+0.5}$ \\
A2107 & & & $428.1/299$ & & $1.3_{-0.3}^{+0.3}$ \\
\noalign{\smallskip}
\hline
\end{tabular}
\end{center}
\end{table}

Cluster masses obtained from the \clmass{} model are given in
Table~\ref{tab:mass}.  Consistent with the model assumptions,
gravitating matter densities are treated as constant within each
shell, so that mass profile within a shell is as given by equation
(\ref{eqn:mass}).  Best fitting chi squareds for the 
unconstrained fits are given in column 2 and the corresponding values
of $M_{500}$ are in column 3.  The best fitting chi squareds for the
monotonic fits are in column 4, with corresponding values of $M_{500}$
and $M_{2500}$ in columns 5 and 6, respectively.  Values used for
$r_{500}$ and $r_{2500}$ are given in Table~\ref{tab:sample}.  There
are no results for $M_{500}$ for Abell~2107 because the \chandra{}
data for it do not reach $\rover$.  No unconstrained result is given
for Abell~1991, the coolest and faintest cluster in the sample,
because the unconstrained \clmass{} model places no useful 90\% upper
limit on $M_{500}$ for it.  Abell~1991 is not detected at high
significance in annuli beyond $\rover$, so that the emission measures
of the corresponding shells can be made arbitrarily small in the
models.  This permits an arbitrarily large gravitating mass in the
shell immediately inside $\rover$.  Making the gravitating matter
density monotonic prohibits the density in the outermost shell within
$\rover$ from exceeding that in the next inner shell, so that the mass
of Abell~1991 can no longer be arbitrarily large.  Thus we do get a
confidence range for $M_{500}$ using the constrained \clmass{} model.

Based on chi squared, all of the fits are formally unacceptable.
However, all clusters show some departures from spherical symmetry and
many also show evidence of ongoing minor merger activity
\citep[\eg][]{mv07}, so that we should expect modest departures from
hydrostatic equilibrium too.  Considered in this light, it is
gratifying that the coarsely resolved \clmass{} model used here gives
reduced chi squareds of less than 2 in most cases.  The clear
exception is Abell~2029, with reduced chi squareds of $\simeq 4.5$.
Abell~2029 is the only cluster in the sample for which there are both
ACIS-I and ACIS-S observations, giving two complete sets of spectra.
Fitting these separately gives reduced chi squareds of 1.4 for ACIS-I
and 2.6 for ACIS-S, more consistent with the other results.  The data
for ACIS-I are a lot shallower than those for ACIS-S (exposures of
$\simeq 10$ ksec and $100$ ksec, resp.), which is why the reduced chi
squared appears better for ACIS-I.  However, the chi squared for the joint
fit is more than a factor of 2 larger than the sum of the chi squareds
for the separate fits, pointing to incompatibility between the two
sets of spectra.  Most likely, this is due to departures from
spherical symmetry in Abell~2029.  Especially on larger scales for the
ACIS-S detectors, the field of view is restricted, so that spectra are
not sampled uniformly from every annulus.  Since the regions sampled
by ACIS-I and ACIS-S differ, anisotropies can lead to significant
differences in the two sets of spectra.  This effect is only evident
when there is more than one set of spectra for a cluster.  It
highlights the imperfections of our assumptions.

\subsection{Mass errors}

To find confidence ranges for the mass, the mass is varied about its
best fitting value and, for each trial mass, a new best fitting model
is determined.  The mass range for which the best fit statistic 
remains below an appropriate limit then defines the confidence range for
the mass.  One virtue of the \clmass{} model is that calculation of
mass confidence ranges can be undertaken as part of the fitting process,
making it possible to marginalize over all free parameters of the fit.

Fixed mass constraints are straightforward to implement.  The total
gravitating mass within the radius $r$ is simply
\begin{equation}
M(r) = \sum_{i=1}^n \rho_i V_i(<r),
\end{equation}
where 
\begin{equation}
V_i(<r) = {4\pi\over3} \cases{r_{i+1}^3 - r_i^3, &for $r_{i+1} < r$\cr
r^3 - r_i^3, &for $r_i < r \le r_{i+1}$\cr 0, &otherwise,\cr}
\end{equation}
is the volume of the \ith{} spherical shell lying inside $r$.  For a
fixed mass, $M(r)$, provided that $V_k(<r) \ne 0$, this is equivalent
to 
\begin{equation} \label{eq:mcons}
\rho_k = {M(r) \over V_k(<r)} - \sum_{i \ne k} \rho_i {V_i(<r) \over
  V_k(<r)},
\end{equation}
which may be used as a constraint on $\rho_k$.  If $\rho_k$ bumps up
against a further constraint (\eg, it goes to zero, or it becomes
non-monotonic when the monotonic condition is applied), that
constraint must take precedence.  The fixed mass constraint can then
be applied to another shell.  In cases where densities of adjacent
shells have to be tied to maintain the monotonic constraint, the fixed
mass constraint can be applied to a group of shells with the same
gravitating mass density (in equation \ref{eq:mcons}, $V_k(<r)$ is
replaced by the total volume inside $r$ for the group of tied shells
and the sum must exclude all shells of that group).  In general,
managing constraints that may conflict with one another is complex.
In an attempt to minimize such conflicts, the fixed mass constraint
was applied to the shell with the largest volume, $V_k(<r)$, since
that choice makes $\rho_k$ least sensitive to $\rho_i$ (under the
constraint, $d\rho_k = - \sum_{i \ne k} d\rho_i V_i(<r) / V_k(<r)$).

Much the same issues apply for determining confidence ranges for the
mass at a fixed overdensity, rather than at a fixed radius.
Specifying the overdensity determines the relationship between $r$ and
$M(r)$ (\ie, $M(r) = 4\pi \rhobar r^3/3$, where the overdensity,
$\rhobar$, is fixed), so that $r$ and $V_i(<r)$ vary as $M$ varies.
However, since $r$ remains fixed for a fixed mass, the remainder of
the discussion above is unaltered.

\section{Discussion}  \label{sec:discussion}

\begin{figure}[t]
\centerline{\includegraphics[width=0.4\textwidth,angle=270]{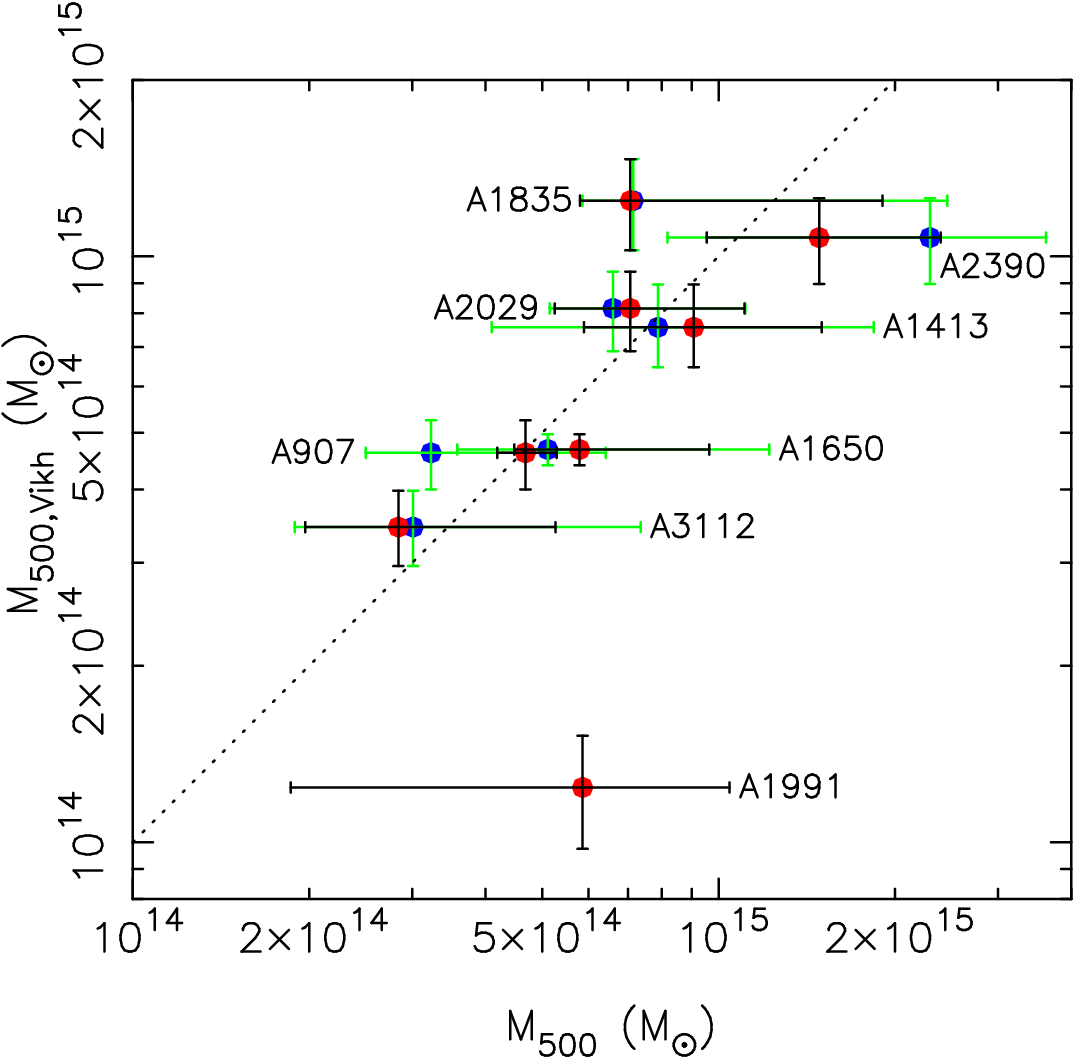}}
\caption{Cluster masses, $M_{500}$, from \citet{vmm05} plotted against
  those from the \clmass{} model.  The unconstrained best fits are
  shown in blue, with their 90\% confidence ranges in green.  With the
  monotonic condition applied, the best fits are shown in red and
  their 90\% confidence ranges in black.  90\% confidence ranges are
  also shown for the masses from \citet{vmm05}.  The dotted line shows
  equality.  No unconstrained result was obtained for Abell
  1991.} \label{fig:m500}
\end{figure}

\subsection{Comparison to conventional mass determinations} \label{sec:massdet}

Fig.~\ref{fig:m500} compares values of $M_{500}$ determined from the
\clmass{} model to those obtained by \citet{vkf06} using essentially
the same spectra.  The figure shows 90\% confidence ranges
($\Delta\chi^2 = 2.706$) for both the case where the gravitating
matter density is unconstrained (best fit blue, confidence range
green) and when it is constrained to be monotonically decreasing (best
fit red, confidence range black).  Plainly, the \clmass{} model does
not place tight limits on values of $M_{500}$ for the sample clusters.
For the unconstrained fits, the 90\% confidence intervals typically
extend over a factor of $\sim 3.5$ in mass.

The situation improves when the monotonic constraint is applied.
Apart from Abell~1991, the average length of the 90\% confidence
ranges is a factor of $\simeq 2.3$ in mass (averaged in log space),
compared to $\simeq 1.34$ (\ie, $\delta M / M \simeq 30\%$) for the
model-dependent results of \cite{vmm05}.  The formal agreement between
the \clmass{} results and \citeauthor{vmm05} is good, apart from
Abell~1991.  The \clmass{} value of $M_{500}$ is high for such a cool
cluster, stemming from a temperature for the outermost shell inside
$\rover$ that is high ($kT \simeq 8\pm4$ keV, $1\sigma$, compared to a
projected temperature of $\simeq 2$ keV).  With a sample of 8
clusters, it is reasonable to expect disagreement at the 90\% level in
one case.  However, the poor result for Abell~1991 mainly reflects
inadequate data at $\rover$ and beyond.  This result highlights the
more general issue for the \clmass{} model, that the total mass is
only well-constrained if we have good X-ray data for annuli inside and
outside the radius of interest.  The lack of better data from beyond
$\rover$ is the primary cause of the large 90\% confidence intervals
for $M_{500}$.

\begin{figure}[t]
\centerline{\includegraphics[width=0.4\textwidth,angle=270]{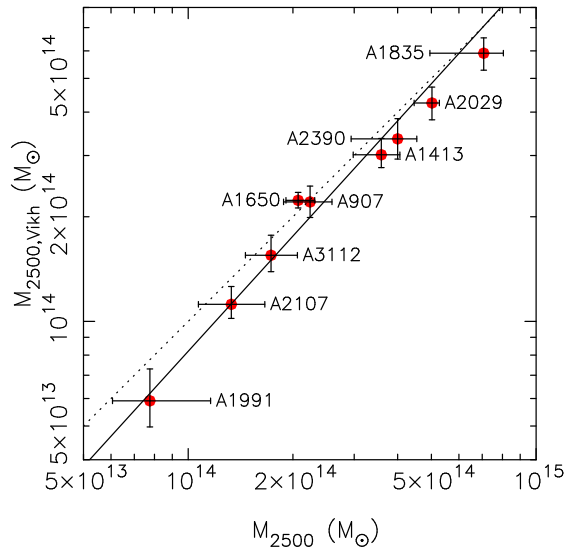}}
\caption{Cluster masses, $M_{2500}$, from \citet{vmm05} versus those
  from the \clmass{} model.  Best fitting results from \clmass{} with
  the monotonic condition and their 90\% confidence ranges are plotted
  in $x$ and results from \citet{vmm05} with their 90\% confidence
  intervals are plotted in $y$. The dotted line shows equality and the
  full line shows the best fitting straight line.} \label{fig:m2500}
\end{figure}

Masses are better determined at smaller radii, where we have good data
from beyond the radius of interest.  For the sample clusters, $\rhover$ is
roughly half of $\rover$ (Table \ref{tab:sample}).  Values of
$M_{2500}$ from monotonic \clmass{} model fits are compared to the
results of \citet{vmm05} in Fig.~\ref{fig:m2500}.  The results for
$M_{2500}$ are markedly more accurate than those for $M_{500}$,
clearly demonstrating the benefit of having high quality spectra that
bracket the radius of interest.  Reflecting the greater freedom of the
model, 90\% confidence ranges from \clmass{} remain larger than those
for the model-dependent masses, with a mean spread for the sample of
$45\%$, compared to $25\%$ for the model-dependent masses.  Individual
values of $M_{2500}$ also agree well with the results of
\citet{vmm05}.

There is a modest offset between the \clmass{} results overall and
those of \citeauthor{vmm05} Using the bisector method of \citet{ab96}
to fit the relationship between the masses
\begin{equation}
\log_{10} {M_{2500,\rm Vikh} \over 10^{14} \msun} 
= a + b \log_{10} {M_{2500} \over 10^{14} \msun}
\end{equation}
gives $a = -0.086$ and $b = 1.101$, which is plotted in
Fig.~\ref{fig:m2500}.  Monte Carlo simulations using the masses from
\citet{vmm05} and the fitted errors from both methods show that $b$ is
within one standard deviation of unity.  However, at the sample mean
(in log space), the fitted relationship gives masses that are $12\%$
low (\ie, the \clmass{} results are high), which occurs by chance
about 2\% of the time.  Thus, there is a moderately significant offset
between the values of $M_{2500}$ obtained by \citeauthor{vmm05} and
those obtained here.  

As noted above, the annuli used here are wider than ideal, so that the
mass profile used by the \clmass{} model is coarsely sampled.  The
gravitational potential, $\Delta\Phi = \int_a^b GM(r)/r^2 \, dr$, is
more sensitive to mass in the inner part of a shell than the outer
part, but, relative to a matter density distribution that decreases
with radius, the constant density approximation used by \clmass{} puts
more mass in the outer part.  Thus, slightly more mass is required in
the coarsely sampled \clmass{} model to obtain the same change in
gravitational potential.  Since the distribution of gas density (\ie,
X-ray emission) is determined by the gravitational potential, that is
the physical quantity which is fixed by fitting the X-ray data.  As a
result, the coarsely sampled \clmass{} model finds masses that are
biased slightly high.  An estimate of this effect is derived in the
Appendix and the bias is found to be no more than $\simeq 3\%$.
Allowing for the bias expected for $\eta = 2$ (see Appendix), the
likelihood of the observed 12\% bias at the mean mass is increased
from about 2\% to $\simeq7\%$.  We note that the bias due to coarse
sampling is quadratic in $(r_{i+1} / r_i - 1)$, so that it decreases
rapidly with finer sampling.  The significance of the remaining
$\simeq 9\%$ mass bias is marginal.  The issue of possible bias in the
\clmass{} model will need to be addressed by other means. 

\subsection{Strengths and Weaknesses} \label{sec:strengths}

The \clmass{} model can approximate arbitrary distributions of
temperature and gravitating matter, but this freedom means that it
fails to constrain cluster masses as tightly as other methods of X-ray
mass determination employing more restrictive assumptions.
Gravitating mass can also move relatively freely between adjacent
shells in the \clmass{} model with little impact on the fit.  Its
performance in this respect can be improved significantly by requiring
the gravitating matter density to be a monotonically decreasing
function of the radius.  This results in smoother mass profiles and
reduces confidence intervals for the mass.  The effect of adding a
relatively weak assumption to the \clmass{} model illustrates the
trade offs between the (apparent) precision of mass determinations and
the assumptions they rely on.

Using some of the best X-ray data available and insisting that the
mass distribution be monotonic, the \clmass{} model still fails to
limit $M_{500}$ to within a factor $\simeq 2$ at the 90\% confidence
level (Fig.~\ref{fig:m500}).  However, it gives well-constrained
results for $M_{2500}$.  The poor results for $M_{500}$ are a
consequence of the generality of the model and the relatively poor
quality of the X-ray data near and beyond $r_{500}$.  This situation
will be slow to improve, because X-ray emission is proportional to the
square of the gas density, making it decline rapidly at large radii.
It suggests that other X-ray mass determinations rely heavily on their
model-dependent assumptions to obtain accurate masses at large radii.
While the issue is most acute for X-ray data, model-dependent
assumptions have the same effect on all mass determinations.  A
model-dependent prior, such as the assumption that a cluster potential
has the NFW form, reduces confidence intervals for its mass.  Unless a
prior is accurate, confidence ranges for the mass that rely on it will
be too optimistic.  Discrepancies between mass determinations may be
caused by inappropriate priors as much as by failures of the physical
assumptions that underlie mass determinations.

Except for clusters that are obviously dynamically active, X-ray mass
determinations generally agree well with dynamical masses
\citep{rgk03} and with masses determined from weak lensing
\citep{mhb08,mrm09}, with a modest bias due to departures from
hydrostatic equilibrium.  Note that the sample of \citet{mhb08}
includes a number of dynamically active clusters, so that their mean
bias of $\sim20\%$ in $M_{500}$ overestimates the bias for relaxed
clusters, such as those used here and by \citet{vkb09}.  Since
model-dependent X-ray masses agree as well as expected with other mass
determinations, the confidence ranges for $M_{500}$ obtained with the
\clmass{} model appear unnecessarily poor.  On the other hand, using
the excellent X-ray data bracketing $\rhover$, the \clmass{} confidence
ranges for $M_{2500}$ are not a lot more conservative than those
obtained by careful modeling of the X-ray data \citep{vmm05}.  Until
there are better X-ray telescopes, use of the \clmass{} model is
probably better restricted to the brighter regions of clusters.  The
results here indicate that model-dependent X-ray determinations of
$M_{500}$ rely, in effect, on extrapolating to $\rover$ models that are
largely constrained by data at smaller radii.  In view of this, using
$M_{2500}$ obtained by fitting the \clmass{} model as a proxy for the
total mass, also an extrapolation in effect, may serve as well.

As noted (section~\ref{sec:prep}), the annuli used here have
$\varpi_{j+1} / \varpi_j = 1.5$, making the gravitating matter density
coarsely sampled for the \clmass{} model.  While spectra with many
photons are needed to obtain accurate temperatures (more so because
\clmass{} effectively deprojects the temperatures), binning the data
coarsely discards information about the radial distribution of surface
brightness that could help to constrain the distribution of
gravitating mass.  The performance of the \clmass{} model may well be
improved by extracting spectra from narrower annuli.  It may even be
advantageous to bin the data so finely that individual spectra alone
have insufficient counts to constrain the temperatures well.  Spectral
parameters for groups of adjacent shells could be tied, in effect
treating them as a single shell for the spectral fits, while
exploiting the information from the individual spectra on the
distribution of brightness.  This and related issues, such as the best
choice of fit statistic, remain to be explored.

There is little constraint on the thermal models that can be used for
the spherical shells with \clmass{}.  In particular, it would be
feasible to use multiphase models \citep[\eg][]{rsf08}, provided that
an appropriately weighted mean temperature is used for the \clmass{}
model temperature for each shell.  If the multiphase gas in a shell is
well-mixed and in local pressure equilibrium at pressure $p$, the
appropriate temperature, $\overline T$, can be determined from the
mean gas density, $\overline \rho_{\rm g}$, by $k \overline T = \mu
\mh p / {\overline\rho_{\rm g}}$.  In terms of the properties of the
separate phases, this yields the weighted temperature $\overline T =
\left( \sum_q K_q T_q^2 \right) / \left( \sum_q K_q T_q \right) $,
where $K_q$ is the \xspec{} norm of the $q^{\rm th}$ phase in the
shell, $T_q$ is its temperature and the sum runs over all gas phases
in the shell.

\section{Conclusions}  \label{sec:conclusions}

A new approach for making X-ray mass determinations of galaxies,
groups and clusters of galaxies was introduced.  The new method relies
on the usual assumptions of hydrostatic equilibrium and spherical
symmetry.  Cluster gas is also assumed to be isothermal and the
gravitating matter density constant in spherical shells centered on a
cluster.  With sufficient shells, this distribution provides a good
approximation to any spherically symmetric distribution of temperature
and gravitating matter, so the approach is largely model-independent.
It does not require the additional, model-dependent assumptions that
are essential to most conventional X-ray mass determinations.

The method has been implemented as an \xspec{} mixing model called
\clmass{}.  With no further constraints, the model only determines the
gravitating mass in individual shells poorly.  This behavior is
improved significantly by constraining the gravitating matter density
to be a monotonically decreasing function of the radius.

The \clmass{} model was used to determine values for $M_{500}$ and
$M_{2500}$ for nine relaxed clusters from the sample of \citet{vmm05},
as far as possible, using the same spectra.  While the results for
$M_{500}$ agree with those of \citeauthor{vmm05}, the \clmass{}
confidence ranges are much larger than those for the more conventional
method.  This reflects the lack of assumptions about the form of the
gravitating matter distribution in \clmass{} and a lack of good X-ray
data from regions beyond $\rover$.  At smaller radii that are bracketed
by good X-ray data, the \clmass{} model performs much better.  For
$M_{2500}$, \clmass{} gives confidence ranges that are only moderately
more relaxed than those of \citet{vmm05}.  Individual masses agree
well between the two methods, but there is a small offset in the
relationship between the two determinations of $M_{2500}$ at the mean
mass.  This is partly a consequence of the coarse radial sampling of
the spectra used here, but there is a marginally significant offset
beyond that.

In regions where we have good X-ray data, the new method can determine
masses with comparable accuracy to conventional methods, without
relying on model-dependent assumptions.  Until we have better X-ray
data for the faint outer regions of clusters, it will be less useful
for determining total cluster masses, other than by using $M_{2500}$,
for example, as a proxy for the virial mass.  In contrast to most
other approaches, the new method allows confidence ranges for the mass
to be determined directly from fits to X-ray spectra.  The method can
be used with a wide range of thermal models for the X-ray emission
from individual shells within a cluster, including multiphase models.



\acknowledgments

This work was partly supported by \chandra\ grant AR6-7016X and NASA
contract NAS8-03060.


Facilities: \facility{CXO}.





\bibliographystyle{apj}

\bibliography{mass}

\appendix

\section{Coarse Sampling Mass Bias}

For the purpose of estimating the mass bias dues to coarse sampling of
the distribution of gravitating matter, we consider a model cluster
in which the density distribution is a power law, $\rho_{\rm t}(r)
\propto r^{-\eta}$, with $\eta < 3$.  Locally, this serves as a
reasonable approximation for more realistic mass distributions.  As
argued in section \ref{sec:massdet}, the gravitational potential
determines the gas distribution, so we assume that the potential is
most constrained by the data.  Thus we need to compare potentials for
the ``true'' cluster and the \clmass{} approximation to it.
Corresponding to the power law density distribution, the mass within
radius $r$ is
\begin{equation}
M_{\rm t}(r) = M_{{\rm t},i} (r/r_i)^{3 - \eta},
\end{equation}
where $M_{{\rm t}, i}$ is the mass within $r_i$.  For this mass
distribution, the change in gravitational potential from $r_i$ to
$r_{i + 1}$ is
\begin{equation} \label{eqn:dpt}
\Delta\Phi_{{\rm t},i} = \int_{r_i}^{r_{i + 1}} {G M_{\rm t}(r) \over r^2}
  \, dr
= {G M_{{\rm t}, i} \over r_i} {(1 + \Delta)^{2 - \eta} - 1 \over 2 -
  \eta},
\end{equation}
where $1 + \Delta = r_{i + 1} / r_i$.  Note that for $\eta \to 2$, the
expression $[(1 + \Delta)^{2 - \eta} - 1] / (2 - \eta) \to \ln (1 +
\Delta)$ is well behaved.

Now consider the \clmass{} model approximation for this cluster.  As
for the sample data, we assume that the shell boundaries follow a
geometric progression, $r_{i + 1} = (1 + \Delta) r_i$, for all $i$,
but we will ignore the cutoff at $i = 1$, assuming that the geometric
progression continues to $i = - \infty$ to simplify the sums (an
accurate approximation).  Matching the power law for the true density
distribution, we also assume that the mass densities for the \clmass{}
model follow the geometric progression
\begin{equation} \label{eqn:geom}
\rho_{i + 1} / \rho_i = (r_{i + 1} / r_i)^{-\eta} = (1 + \Delta)^{-\eta}.
\end{equation}
This condition is not strictly required, but, like the power law for
the ``true'' density, it should be a reasonable local approximation.
The form (\ref{eqn:geom}) determines the ratio $\Delta M_i / M_i$
found below (\ref{eqn:biasmass}), making it critical to our
estimate of the mass bias.  For the \clmass{} model, the mass in the
\ith{} shell is then
\begin{equation} \label{eqn:dmbias}
\Delta M_i = {4 \pi \over 3} \rho_i (r_{i+1}^3 - r_i^3)
=  {4 \pi \over 3} \rho_i r_i^3 [(1 + \Delta)^3 - 1].
\end{equation}
Under our assumptions, we have $\Delta M_k = \Delta M_i (1 +
\Delta)^{(k - i)(3 - \eta)}$, so that the shell masses form a
geometric progression.  Thus, the total mass within $r_i$ for the
\clmass{} model can be summed to give the total mass within $r_i$,
\begin{equation} \label{eqn:biasmass}
M_i = \sum_{k = -\infty}^{i - 1} \Delta M_k = {\Delta M_i \over (1 +
  \Delta)^{3 - \eta} - 1}.
\end{equation}
The increase in potential in the \ith{} shell is (equation
\ref{eqn:dphi})
\begin{equation}
\Delta\Phi_i = \left[ {G M_i \over r_i} + {2 \pi \over 3} G \rho_i
  r_i^2 \Delta (3 + \Delta) \right] {\Delta \over 1 + \Delta}
\end{equation}
and using (\ref{eqn:dmbias}) and (\ref{eqn:biasmass}) to express $G
\rho_i r_i^2$ in terms of $GM_i/r_i$, this becomes
\begin{equation} \label{eqn:dpcm}
\Delta\Phi_i = {G M_i \over r_i} \left[ 1 + {\Delta (3 + \Delta) \over
    2} {(1 + \Delta)^{3 - \eta} - 1 \over (1 + \Delta)^3 - 1} \right]
          {\Delta \over 1 + \Delta}. 
\end{equation}

If the potential is determined exactly by the data, as argued above,
then fitting the \clmass{} model should make $\Delta \Phi_i = \Delta 
\Phi_{{\rm t},i}$.  From equations (\ref{eqn:dpt}) and
(\ref{eqn:dpcm}), this determines the ratio of the true mass to the
mass obtained from the \clmass{} model as 
\begin{equation}
{M_{{\rm t}, i} \over M_i} = 1 - b
= {2 - \eta \over (1 + \Delta)^{2 - \eta} - 1}
\left[1 + {\Delta (3 + \Delta) \over 2} {(1 + \Delta)^{3 - \eta} - 1
    \over (1 + \Delta)^3 - 1} \right] {\Delta \over 1 + \Delta} 
\simeq 1 - {\eta (3 - \eta) \over 12} \Delta^2,
\end{equation}
where $b$ is the mass bias.  The last form is an approximation
showing the lowest order deviation from unity.  The \clmass{} model
makes the matter density uniform in shells, so there is no bias for a
uniform density distribution, \ie, for $\eta = 0$.  The bias is also
zero for $\eta \to 3$, since then the mass is concentrated at the
cluster center and $\Delta\Phi_i$ is dominated by the potential due to
mass inside $r_i$, which is determined correctly in \clmass{}.  For
the cluster sample used here, $\Delta = r_{i+1} / r_i - 1 = 0.5$.  The
bias for $\Delta = 0.5$ is overestimated somewhat ($\simeq 60\%$) by
the approximation $b \simeq \eta (3 - \eta) \Delta^2 / 12$.
Nevertheless, $b(\Delta = 0.5)$ is maximized for $\eta = 1.5$ at $b
\simeq 0.030$.  Observed values for $\eta$ at $\rhover$ are about 2 or
a little greater.  For $\eta = 2$, the model described here gives $b
\simeq 0.026$.

\end{document}